\documentclass[conference]{IEEEtran}
\IEEEoverridecommandlockouts
\usepackage{float}
\usepackage[caption = false]{subfig}
\usepackage{spconf,amsmath,graphicx}
\usepackage{xcolor}
\usepackage{hyperref}
\usepackage{lipsum,multicol}
\usepackage{amsfonts}
\usepackage{ulem}
\usepackage[letterpaper, left=0.75in, right=0.60in, bottom=0.60in, top=1in]{geometry}
\def\BibTeX{{\rm B\kern-.05em{\sc i\kern-.025em b}\kern-.08em
    T\kern-.1667em\lower.7ex\hbox{E}\kern-.125emX}}
\begin{document}

\graphicspath{ {img/} }
\def\x{{\mathbf x}}
\def\L{{\cal L}}


\title{Image storage on synthetic DNA using  compressive autoencoders and DNA-adapted entropy coders}
\name{Xavier Pic$^{\star}$, Eva Gil San Antonio$^{\star}$, Melpomeni Dimopoulou$^{\star}$,  Marc Antonini$^{\star}$}
\address{$^{\star}$ I3S laboratory, Côte d’Azur University and CNRS - UMR 7271\\2000 Rte des Lucioles, Les Algorithmes, Euclide B, Sophia Antipolis, France}

\maketitle

\begin{abstract}
\par

Over the past years, the ever-growing trend on data storage demand, more specifically for "cold" data (rarely accessed data), has motivated research for alternative systems of data storage.
Because of its biochemical characteristics, synthetic DNA molecules are now considered as serious candidates for this new kind of storage. This paper presents some results on lossy image compression methods based on convolutional autoencoders adapted to DNA data storage, with synthetic DNA-adapted entropic and fixed-length codes. 

The model architectures presented here have been designed to efficiently compress images, 
encode them into a quaternary code, and finally store them into synthetic DNA molecules. 
This work also aims at making the compression models better fit the problematics that we encounter when storing data into DNA, 
namely the fact that the DNA writing, storing and reading methods are error prone processes. 
The main take aways of this kind of compressive autoencoder are our latent space quantization and the different DNA adapted entropy coders used to encode the quantized latent space, which are an improvement over the fixed length DNA adapted coders that were previously used.
\end{abstract}
%
\vspace{-0.5\baselineskip}
\section{Introduction}
\vspace{-0.2\baselineskip}
\label{sec:intro}
\par
The memory of humanity relies on our ability to manage increasingly large amounts of data, over periods of time ranging from a few years to several centuries. Current tools are no longer sufficient and it is necessary to consider game-changing solutions that can become operational quickly. One of the most promising solutions is to store information in the form of DNA, just like the genome by living beings. Indeed, DNA provides a very stable storage medium over very long periods with simple implementation conditions \cite{DNARAM}.
\par
The first step in the encoding workflow is the construction of a dictionary of codewords composed by the symbols A, G, C and T also called {\it nucleotides (nts in short)}. The DNA coded information stream must respect some biochemical constraints on the combinations of bases that form a DNA fragment: homopolymers, high/low GC content and repeated patterns should be avoided. One must also take into account that the process involves some biochemical procedures which can corrupt the stored data. Synthesis, sequencing, storage and the manipulation of DNA (mainly PCR amplification) may introduce errors by introducing substitutions and indels (insertions or deletions of nts), and may jeopardize the integrity of the stored content \cite{Goldman2013}.
\par
Autoencoders have become a focus in the field of image compression. Quite early, some autoencoders \cite{Theis},\cite{Balle} and recurrent neural networks \cite{Toderici} were already showing better performances than the more conventional compression algorithms like JPEG or JPEG2000. Some research \cite{Fei},\cite{Yoojin} aimed at having variable bit-rate compression autoencoders, other works put their focus on the differentiability problem of the quantization \cite{Theis},\cite{dumas2018autoencoder} or the entropy evaluation \cite{Theis},\cite{oliveira}. Our goal is to adapt compression autoencoders to DNA data storage. For that, we have based our proposed network on the compression autoencoder designed by Lucas Theis et al. in \cite{Theis}, and introduced a DNA coder for the autoencoder's latent space tensor.
\vspace{-0.5\baselineskip}
\section{Context}
\vspace{-0.2\baselineskip}
\subsection{DNA data storage}
\vspace{-0.2\baselineskip}
\par 

The first application of DNA data storage took place in 2012 with the work presented by Church and his team in \cite{church2012next}. It consisted on a simple encoding for binary to quaternary conversion, storing a 659-Kbyte book into DNA. This laid the groundwork for understanding the constraints and errors in DNA encoding. Later works introduced error correction mechanisms to minimize sequencing errors. In 2013, Goldman et al. proposed the first constrained encoding algorithm avoiding homopolymers and employing redundancy for error correction. \cite{Goldman2013}. Grass et al. were the first to incorporate Reed Solomon codes for error correction in 2015 \cite{grass2015robust}, and in 2016, Blawat et al. proposed creating multiple dictionaries for each input symbol, facilitating forward error correction. They also introduced Reed Solomon codes to strengthen the headers and ensure reliable decoding \cite{blawat2016forward}. In the same year, Erlich et al. presented a method based on Fountain codes, creating multiple oligos and discarding those violating biochemical constraints \cite{erlich2017dna}. Also in 2016, Borhholt et al. presented a DNA-based archiving system as part of a Microsoft research that integrated random access \cite{bornholt2016dna}. In 2019, a Microsoft team demonstrated a fully automated end-to-end system \cite{takahashi2019demonstration}.

However, to ensure better adaptation to the characteristics of the storage medium, i.e., DNA, and possibly achieve higher storage efficiencies, it is better to design coding algorithms specific for DNA storage. Indeed, since DNA synthesis cost is relatively high, it is also important to take full advantage of an optimal compression that can be achieved before synthesizing the sequence into DNA.


\par
\subsection{Image coding solutions for DNA data storage}
\vspace{-0.2\baselineskip}

Early DNA data storage methods were mostly general solutions for any digital data, often compressing images with JPEG before encoding and translating into quaternary. However, as the field evolved, encoding adapted to the source. 

In 2019, Dimopoulou et al. proposed one of the first works introducing image compression using Discrete Wavelet Transform and scalar quantization and then encoded using a constrained quaternary code \cite{dimopoulou2019biologically}. The algorithm was later improved with vector quantization \cite{dimopoulou2021imageVQ}. Later in 2021, the same team presented a variable-length solution based on the JPEG standard for image compression, replacing the original binary encodings with quaternary ones \cite{dimopoulou2021jpeg}. On the same year, Pan et al. proposed to store quantized images into DNA, integrating Huffman coding and post processing techniques based on inpainting to correct discolorations on the decoded image \cite{pan2019image}. 
In 2021, we introduced a novel entropy coder \cite{SFC4} for DNA data storage that had increased performance over the state of the art Huffman/Goldman DNA entropy coder \cite{Goldman2013}. This novel entropy coder was then introduced in a JPEG-based image codec with increased performance over the state of the art JPEG-based image codec \cite{dimopoulou2021jpeg}.
Also in 2021, motivated by the large amount if images already stored in JPEG format,  Secilmis et al. introduced a JPEG transcoder specifically designed and adapted to encode JPEG-compressed binary data as an attempt to ease the migration from digitl storage systems to DNA-based systems \cite{secilmis2022towards}.

More recently, in 2022 in \cite{pic2023rotating}, we developed a new entropy coder for block-based image-coding schemas. This coder introduces variability in quaternary streams and reduces homopolymers and repeated patterns to decrease error likelihood, while not affecting the compression rate. The coder was integrated into four existing JPEG-inspired DNA coders, demonstrating improved robustness and biochemical constraint compliance.

Following the example of what is done today in the context of JPEG standardization for image coding with the development of JPEG AI\footnote{\color{blue}\url{https://jpeg.org/jpegai/index.html}}, we propose in this work to use neural networks such as autoencoders for image coding on quaternary codes.

\vspace{-0.5\baselineskip}
\subsection{Compressive autoencoders for DNA image storage}
\vspace{-0.2\baselineskip}

\begin{figure*}[!ht]
    \centering
    \includegraphics[scale=0.5]{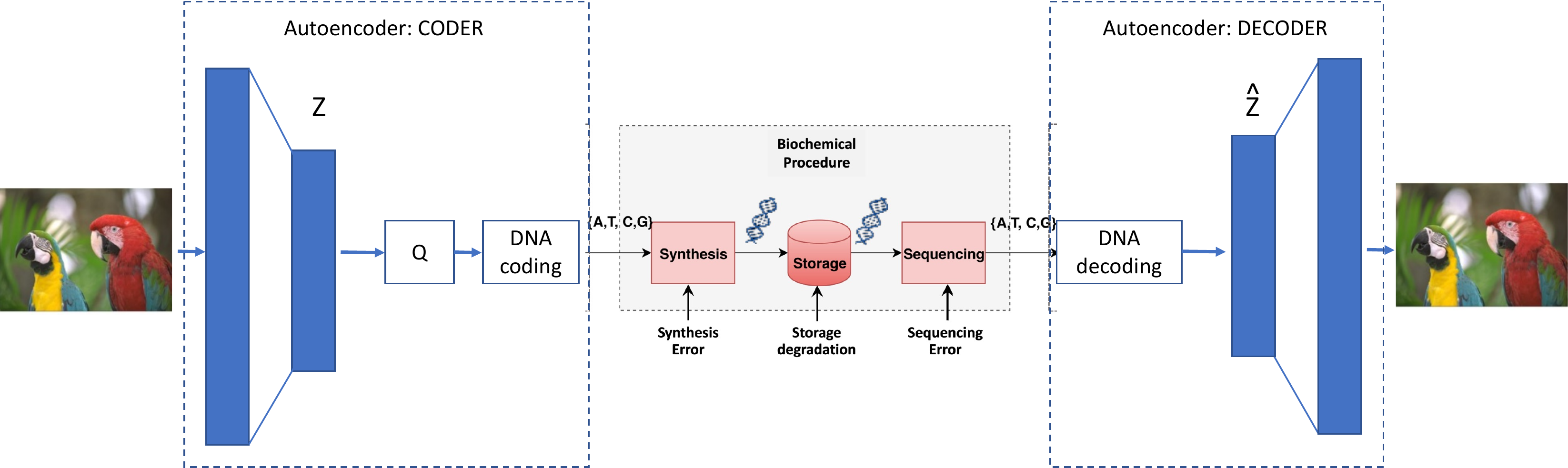}
    \vspace{-0.5\baselineskip}
    \caption{General scheme of an autoencoder used for image storage in synthetic DNA. $Z$ is the latent vector in the latent space. For compression purposes, $Z$ is quantized and encoded in a quaternary code for DNA storage. Then the quaternary code is used for DNA synthesis (writing). The image reconstruction process needs  (reading) before decoding with the autoencoder.}
    \label{fig:Auto}
\vspace{-1\baselineskip}
\end{figure*}

The general idea of our proposed encoding process is depicted in Fig. \ref{fig:Auto} and can be very roughly described by the following steps. Firstly, the input image has to be compressed using a lossy/near
lossless image coder. Here, we propose to use a compressive autoencoder to learn the image characteristics as well as the biochemical noisy process. Convolutionnal autoencoders have shown good behaviour for image denoising and could then appear as a good solution when it comes to deal with the noise introduced by the DNA data storage process. Secondly, a quantization operation is introduced to quantize the latent vector $Z$ in the latent space. The reason for that is that the latent space does not suffice as a compressed representation because it uses floating numbers. The goal of the quantization is to reduce the number of possible values that would be encoded, hence reducing the coding cost. Thirdly, in the case of DNA storage, the latent vector is encoded using a quaternary code using the $\{A,C,T,G\}$ alphabet. Decoding is ensured thanks to the decoder.
The quantization operation is one of the major problems when training compressive autoencoders. Indeed, to be able to train a model, all the operations have to be differentiable, which is not the case for the quantization (most of the times it is equivalent to a rounding function). However, this can be tackled by the use of a linear approximation function as in \cite{Theis}.
\par
During training, the compressive autoencoder has to minimize two quantities that will be mathematically defined in next section:
\begin{itemize}
    \item The distortion which corresponds to the squared difference between the original image and the image reconstructed after compression and decompression.
    \item The entropy of the quantized latent space computed in a quaternary basis. The entropy being closely linked to the rate, we will use indifferently both terms in the rest of the paper.
\end{itemize} 
The structure of the compressive autoencoder model we proposed in this paper is closely linked to the one proposed by Theis et al. in \cite{Theis}. It is described in the following section \ref{non-noisy}.

\par
\par
\vspace{-0.3\baselineskip}
\section{Encoding}
\label{non-noisy}
\vspace{-0.3\baselineskip}
\subsection{Neural Network}
\vspace{-0.2\baselineskip}
\par
The proposed autoencoder was designed to obtain a latent space with higher dimensionality than the one proposed in \cite{Theis}, the objective of such a model was to obtain higher bit-rates.
This means that our new model now has a reduced number of parameters, which could cause some overfitting problems during the training process.
To compensate such a risk, we decided to add new residual blocks to the model with skipped connections.
\par
When encoding data, the common practice is to operate with integers, not with floating point numbers. For that reason, the output $Z$ of an autoencoder group of layers in the latent space cannot be encoded using quaternary code as it is and a quantization operator must then be introduced in the process. Given an input image $I$, the output image (decoded image) $\hat{I}$ of the autoencoder can be expressed as a combination of several functions such as follows:
\begin{equation}
    \hat{I} = g(\beta_{DNA}(\alpha_{DNA}(Q(f(I))))
\label{autofunc}
\end{equation}
where, $f$ represents the encoding part of the autoencoder and $g$ the decoding, $Q$ represents the uniform quantizer of step $q$ that rounds the components z of the latent vector $Z=f(I)$ into integer values and is defined as:
\begin{equation}
\label{Quantizer}
    Q(z)=q\times\left \lfloor\frac{z}{q}+\frac{1}{2}\right \rfloor
\end{equation}
Finally, $\alpha_{DNA}$ and $\beta_{DNA}$ represent the DNA-encoding and decoding algorithms that encode a sequence of symbols into a quaternary stream composed by the letters of the alphabet $\{A,C,T,G\}$. When no noise is introduced, it is clear that $\beta_{DNA}(\alpha_{DNA}(k))=k \ \forall k \in \mathbb{Z}$. In this work we have used the DNA fixed-length code proposed in \cite{DNAcoding}.
\vspace{-0.3\baselineskip}
\subsection{Loss function}
\vspace{-0.2\baselineskip}
In this work, we used the following classical loss function:
\begin{equation}
\label{Loss}
    L = ||I-\hat{I}||^2 + \lambda H(Q(Z)),
\end{equation}
where $I$ and $\hat{I}$ are respectively the input and output images and $H(Q(Z))$ corresponds to the entropy of the quantized latent space computed in base $4$ as follows:
\begin{equation}
    H(Q(Z))= -\sum_{i=1}^n Pr\{Q(z_i)\}log_4 Pr\{Q(z_i)\}
\end{equation}
with $Q(Z)=(Q(z_1),Q(z_2),...,Q(z_n))$ and $Pr\{Q(z_i)\}$ the probability of a quantized component. The entropy $H$ is expressed in {\it nucleotides per component}.
\par
The compression model output was bounded using a hyperbolic tangent function.
The uniform quantization was then applied to the output of that bounded function, giving a finite number of different possible values.
The number of possible values can be adjusted with the quantization step $q$.
\vspace{-0.3\baselineskip}
\subsection{Encoding the latent space}
\vspace{-0.2\baselineskip}
Since the Neural Network is designed to minimize the entropy of the latent space, it makes sense to use a DNA-like entropy coder to encode the latent space quantized values. In the case where we assume the DNA data storage channel to be noiseless, an entropy coder is a promising candidate because of its increased compression performance.
We identified two possible coders that can be used for the latent space: the Huffman/Goldman coder \cite{Goldman2013} and the SFC4 coder that we presented in \cite{SFC4}.
\par
\begin{figure}
    \centering
    \includegraphics[scale=0.24]{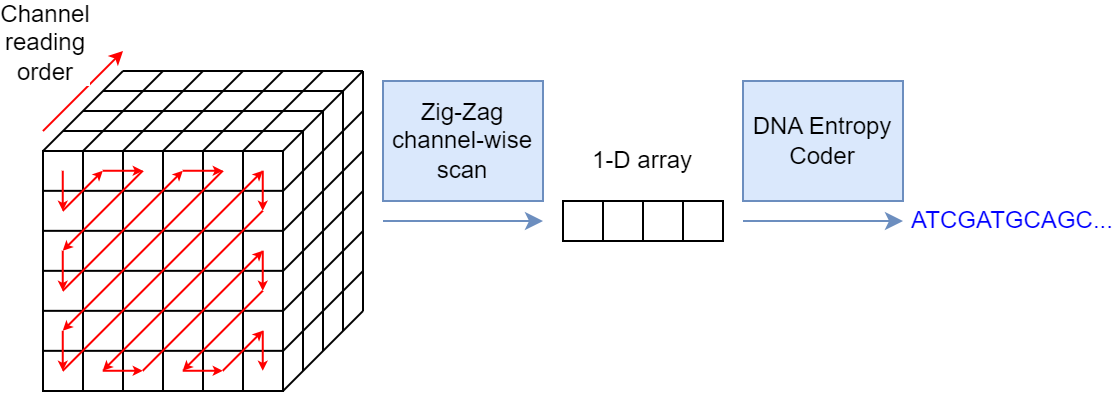}
\vspace{-1.5\baselineskip}
    \caption{Coding scheme of the autoencoder's latent space}
    \label{fig:coding_scheme}
\vspace{-1.5\baselineskip}
\end{figure}

The tensor of quantized coefficients is first transformed into a one dimensional array of coefficients with a two dimensional zig-zag scan of each channel of the tensor as presented in Fig. \ref{fig:coding_scheme}. The coefficients of this one dimensional array are encoded into a quaternary stream with the help of one of the entropy coders previously described. The statistics of the latent space $Q(Z)$ are computed to instantiate the entropy coder: for each symbol $\alpha$ (quantization level) found in the source, the number of occurrences $N(\alpha)$ is computed. Minimizing $H(Q(Z))$ during training will help improve the performance of our entropy coder since the occurrences $N(\alpha)$ of each symbol $\alpha$ and the general entropy $H(Q(Z))$ of the quantized latent space are highly correlated, and because entropy coders perform well on low entropy sources.

When decoding, the one dimensional array is first retrieved thanks to the entropy coder.
To reconstruct the tensor from this one dimensional array, we first create a tensor of same dimensions containing only zeros, and we then fill this tensor with the coefficients coming from the one-dimensional array, in the order defined by the previously described scan. This requires the decoder to know the dimensions of the tensor. This information has to be transmitted.
\vspace{-0.3\baselineskip}
\subsection{Formatting}
\vspace{-0.2\baselineskip}
Due to the current restrictions of DNA synthesis technology, the length of the constructed strands is limited to up to 300 nts. Hence, the quaternary stream has to be cut into smaller chunks called oligos. For this work, we decided to use a length of 200 nts, following the default values used in the JPEG DNA Benchmark Codec and other works \cite{dimopoulou2021jpeg, SFC4}. After synthesis, all the oligos will be suspended in the same solution, and mixed with other stored files. This means that when someone will retrieve the data, the oligos will be unordered and not indexed. So to ensure the decodability of the stored data, the oligos belonging to a file will first need a file-specific identifier. This identifier will be barcoded to be able to chemically retrieve the oligos relating specifically to this file, without sequencing the rest. The oligos pertaining to a file also have to be re-ordered correctly to ensure decodability. To this end, an index is included in each oligo of the encoded latent space, describing the position of the oligo in the pool of oligos encoding this latent space. This index is a number encoded in quaternary using a fixed length coder. The oligos resulting from this formatting process are called Data Oligos (DO) as shown in Fig. \ref{fig:formatting}.
\vspace{-0.3\baselineskip}
\subsection{Transmission of frequency tables}
\vspace{-0.2\baselineskip}
In the case where a variable length coder is used, like the Huffman/Goldman or the SFC4 coders, additional information must be transmitted to be able to use the same instance of these coders. This is due to the fact that, depending on the statistic of the source, these coders will not use the same set of codewords. The decodability of the whole data lies on the assertion that both the encoder and the decoder use the same instance of the variable length coder. For this reason, transmitting the information relating to this instance of the variable length coder is crucial. The first solution would be to directly transmit the set of codewords, but given the high error rate of our channel, if one of the codewords is modified by an error, the propagation of this error would be catastrophic. Instead, we decided to transmit the statistics of the source. In other words, for each symbol $\alpha$ on the alphabet $\Sigma$ of the source $S$, we compute the number of occurrences of this symbol in the source. We then encode the array of occurrences of all the symbols of the source into a quaternary stream, and format the stream into oligos, with the same barcode file identifier used for the rest of the data. We also added a special identifier stating that these oligos described the statistics of the source. Finally, in the same manner as for the first oligos, an index is encoded to reorder the quaternary stream of the coded frequencies. The oligos describing the source statistics are called Frequencies Information Oligos (FIO) and are shown in Fig. \ref{fig:formatting}.
\vspace{-0.3\baselineskip}
\subsection{Transmission of metadata}
\begin{figure*}
    \centering
    \includegraphics[scale=0.25]{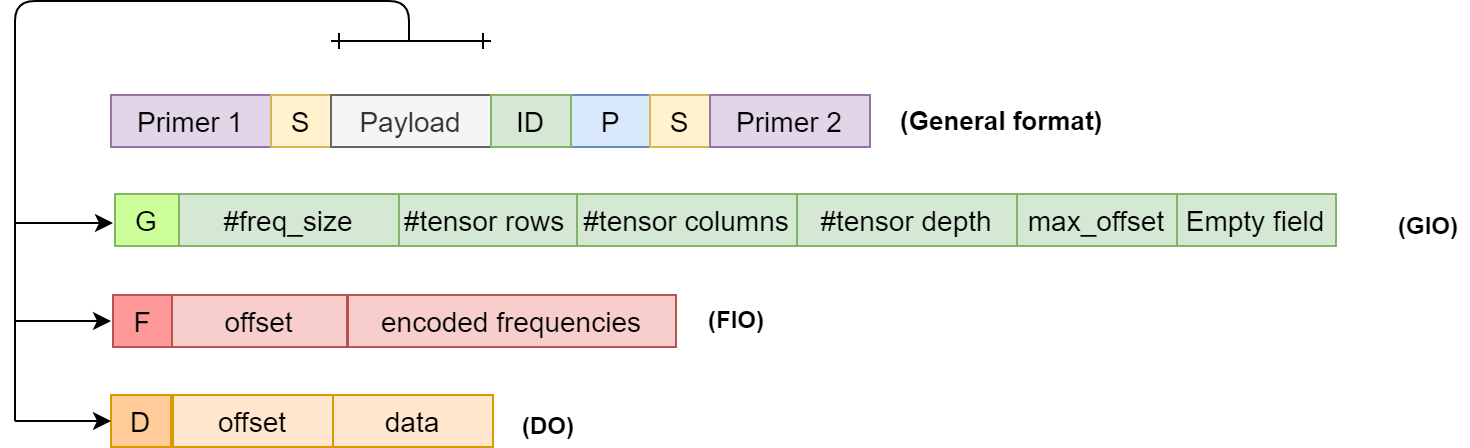}
    \vspace{-0.25cm}
    \caption{Formatting of the oligos for the Compressive Autoencoder. In Green, the General Information Oligo (GIO) describes the shape of the tensor and some parameters of the codec. In red, the Frequencies Information Oligos (FIO) describe the statistics of the source, and in orange, the Data Oligos (DO) encode the latent space.}
    \label{fig:formatting}
    \vspace{-0.4cm}
\end{figure*}

\vspace{-0.2\baselineskip}
The last information that need to be transmitted is packed into a single oligo, called the General Information Oligo (GIO), see Fig. \ref{fig:formatting}. It has the same file-specific identifier as the other oligos of the same file, and identifier stating that this oligo is the GIO. It contains information on how the FIOs where encoded, and on the shape of the encoded tensor. Without this knowledge of the tensor shape, it would be impossible to retrieve the correct tensor from a one dimensional stream.
\vspace{-0.4\baselineskip}
\section{Experimental results}
\vspace{-0.3\baselineskip}
\subsection{Implementation of the training}
\vspace{-0.2\baselineskip}
We trained the model with the 30k Flickr image dataset. 
During the training step, we used batches of 32 random crops of size 96x96 from the Flickr images. 
The training process has been separated into two steps: the first with a learning rate of 1e-4 during 200 epochs and the second one of 1e-5 that would be used for 500 epochs.
Models have been trained independently for each quantization step.
\par
 After training our novel networks as described in section \ref{non-noisy} using the model of formula (\ref{autofunc}), we evaluated their performance. The experiments were conducted on different autoencoder models, each one trained to a given compression rate (given quantization step $q$ as defined in formula (\ref{Quantizer})).

\vspace{-0.3\baselineskip}
\subsection{Evaluation process}
\vspace{-0.2\baselineskip}
The models were evaluated using the Kodak dataset\footnote{\color{blue}\url{http://r0k.us/graphics/kodak/}}. For each chosen compression rate, we computed the performances for each image of the dataset, and also the average on the whole dataset. 

 The final curves show the evaluation of the PSNR in function of the compression rate $C_r$ in bits/nt.
 \begin{equation}
 C_r=\frac{H\times W\times d}{N_{nuc}}
 \end{equation}
 $H$, $W$ and $d$ are the height, width and pixel depth of the input image respectively, and $N_{nuc}$ the number of nts in the encoded oligo pool.

To evaluate the performance of the autoencoder, both Hufmman/Goldman \cite{Goldman2013} and our Shannon Fano Constrained entropy coder \cite{SFC4} have been tested to encode the latent space.
The performance of our novel codec was compared with other state of the art image codecs adapted to synthetic DNA data storage. It was also compared to the same compressive autoencoder, but with a tensor encoded by a fixed length constrained coder. 

\vspace{-0.4\baselineskip}
\subsection{Results}
\begin{figure*}
    \centering
    \includegraphics[scale=0.5]{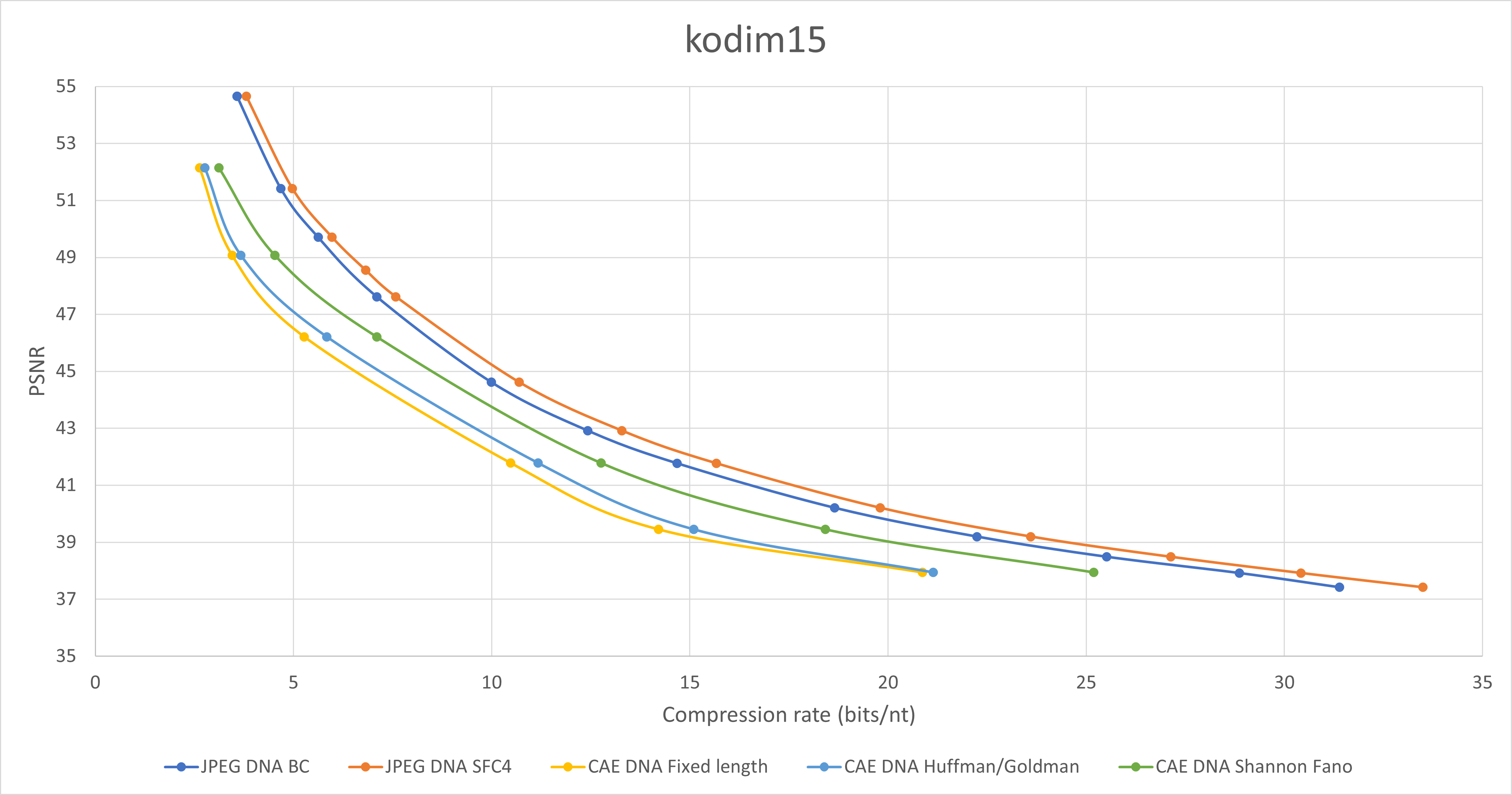}
    \vspace{-0.25cm}
    \caption{Rate distortion curves for the image kodim15 from the Kodak dataset}
    \vspace{-0.4cm}
    \label{fig:kodim15}
\end{figure*}

\begin{figure*}
    \centering
    \includegraphics[scale=0.5]{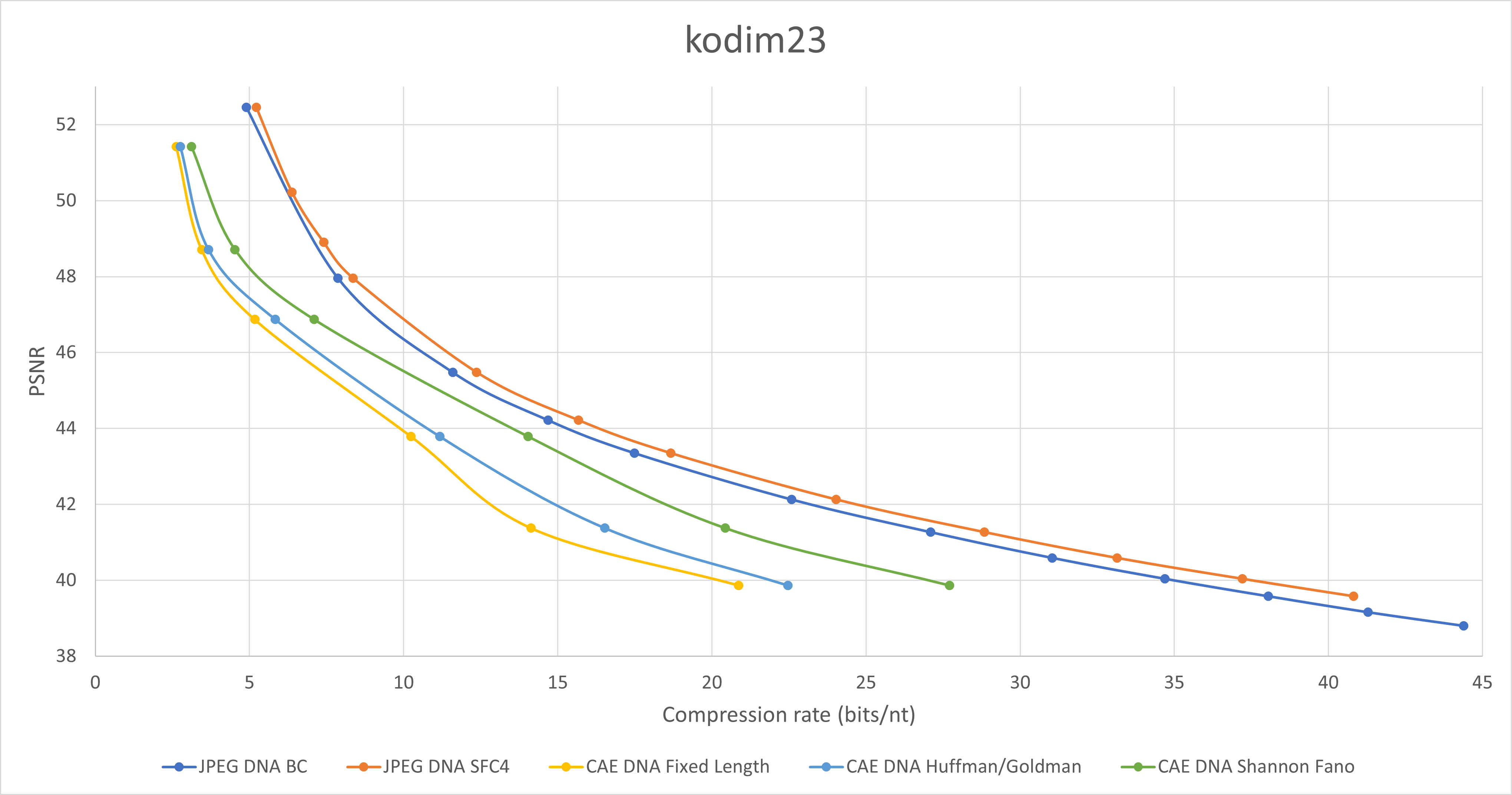}
    \vspace{-0.25cm}
    \caption{Rate distortion curves for the image kodim23 from the Kodak dataset}
    \label{fig:kodim23}
\end{figure*}

\vspace{-0.2\baselineskip}
Figures \ref{fig:kodim15} and \ref{fig:kodim23} show examples of performance over two of the images of the Kodak dataset.
The autoencoder-based methods (CAE DNA Fixed Length, CAE DNA Huffman/Goldman and CAE DNA Shannon Fano in the figures) still underperform in comparison to the JPEG-based ones (JPEG DNA BC and JPEG DNA SFC4), but the entropy coders have improved the compression rate of the autoencoder based methods in comparison to the fixed length coder, especially Shannon Fano. The rate distortion curves computed on the rest of the Kodak dataset are consistent with the ones shown here.

\vspace{-0.3\baselineskip}
\section{Conclusion}
\vspace{-0.2\baselineskip}
In this work, we have developed a compression solution for image storage on synthetic DNA, robust to substitution noise. The proposed approach is based on compressive autoencoder.
The latent space of this autoencoder is quantized and encoded into DNA, with an entropy coder adapted to DNA data storage. These entropy coders (Huffman/Goldman and Shannon Fano) have been compared in performance in (Figures \ref{fig:kodim15} and \ref{fig:kodim23}) to methods where the same latent space is encoded using a fixed-length coder (CAE Fixed Length) and to JPEG-based methods (JPEG DNA BC, JPEG DNA SFC4). The gain in compression rate with entropy coders is significant, especially with the Shannon Fano based entropy coder, although the JPEG-methods still perform better. To improve the performance of the autoencoder based compression methods and given that the latent space is composed of large zones of quantized values that are the same, a run-length based coding could boost the coding performance.
\par
In future works, since the processes involved in DNA data storage are very error prone, a study on the noise robustness of these novel methods could be explored. The introduction of error correction codes could counterbalance the effect of those errors, at the cost of a loss in rate. The autoencoder could also be trained to adapt the compression to this noisy channel. More performant autoencoders like the JPEG AI compression model could also be experimented to improve the rate. 
\vspace{-0.5\baselineskip}
\bibliographystyle{IEEEbib}
\bibliography{strings}

\begin{thebibliography}{10}

\bibitem{DNARAM}
S.~M. Hossein~Tabatabaei Yazdi, Ryan Gabrys, and Olgica Milenkovic,
\newblock ``Portable and error-free dna-based data storage,''
\newblock {\em Nature}, 2017.

\bibitem{Goldman2013}
Nick Goldman, Paul Bertone, Siyuan Chen, Christophe Dessimoz, Emily~M LeProust,
  Botond Sipos, and Ewan Birney,
\newblock ``Towards practical, high-capacity, low-maintenance information
  storage in synthesized {DNA},''
\newblock {\em Nature}, vol. 494, no. 7435, pp. 77, 2013.

\bibitem{Theis}
Lucas Theis, Whenzhe Shi, Andrew Cunningham, and Ferenc Huszár,
\newblock ``Lossy image compression with compressive autoencoders,''
\newblock {\em International Conference on Learning Representations}, 2017.

\bibitem{Balle}
Ballé Johannes, David Minnen, Singh Saurabh, Sun~Jin Hwang, and Nick Johnston,
\newblock ``Variational image compression with a scale hyperprior,''
\newblock {\em International Conference on Learning Representations}, 2018.

\bibitem{Toderici}
George Toderici, Damien Vincent, Nick Johnston, Sun~Jin Hwang, David Minnen,
  Joel Shor, and Michelle Covell,
\newblock ``Full resolution image compression with recurrent neural networks,''
\newblock {\em IEEE Conference on Computer Vision and Pattern Recognition},
  2017.

\bibitem{Fei}
Fei Yang, Luis Herranz, Joost van~de Weijer, José A.~Iglesias Guitián,
  Antonio~M. López, and Mikhail~G. Mozerov,
\newblock ``Variable rate deep image compression with modulated autoencoder,''
\newblock {\em IEEE Signal Processing Letters}, vol. 27, pp. 331--335, 2020.

\bibitem{Yoojin}
Yoojin Choi, Mostafa El-Khamy, and Jungwon Lee,
\newblock ``Variable rate deep image compression with a conditional
  autoencoder,''
\newblock {\em IEEE/CVF International Conference on Computer Vision (ICCV)},
  2019.

\bibitem{dumas2018autoencoder}
Thierry Dumas, Aline Roumy, and Christine Guillemot,
\newblock ``Autoencoder based image compression: can the learning be
  quantization independent?,''
\newblock 2018.

\bibitem{oliveira}
V~Oliveira, T~Oberlin, M.~Chabert, Charly Poulliat, Mickaël Bruno, C~Latry,
  M~Carlavan, S~Henrot, F~Falzon, and Roberto Camarero,
\newblock ``Simplified entropy model for reduced-complexity end-to-end
  variational autoencoder with application to on-board satellite image
  compression,''
\newblock 09 2020.

\bibitem{church2012next}
George~M Church, Yuan Gao, and Sriram Kosuri,
\newblock ``Next-generation digital information storage in {DNA},''
\newblock {\em Science}, p. 1226355, 2012.

\bibitem{grass2015robust}
Robert~N Grass, Reinhard Heckel, Michela Puddu, Daniela Paunescu, and
  Wendelin~J Stark,
\newblock ``Robust chemical preservation of digital information on {DNA} in
  silica with error-correcting codes,''
\newblock {\em Angewandte Chemie International Edition}, vol. 54, no. 8, pp.
  2552--2555, 2015.

\bibitem{blawat2016forward}
Meinolf Blawat, Klaus Gaedke, Ingo Huetter, Xiao-Ming Chen, Brian Turczyk,
  Samuel Inverso, Benjamin~W Pruitt, and George~M Church,
\newblock ``Forward error correction for {DNA} data storage,''
\newblock {\em Procedia Computer Science}, vol. 80, pp. 1011--1022, 2016.

\bibitem{erlich2017dna}
Yaniv Erlich and Dina Zielinski,
\newblock ``{DNA} fountain enables a robust and efficient storage
  architecture,''
\newblock {\em Science}, vol. 355, no. 6328, pp. 950--954, 2017.

\bibitem{bornholt2016dna}
James Bornholt, Randolph Lopez, Douglas~M Carmean, Luis Ceze, Georg Seelig, and
  Karin Strauss,
\newblock ``A {DNA}-based archival storage system,''
\newblock {\em ACM SIGOPS Operating Systems Review}, vol. 50, no. 2, pp.
  637--649, 2016.

\bibitem{takahashi2019demonstration}
Christopher~N Takahashi, Bichlien~H Nguyen, Karin Strauss, and Luis Ceze,
\newblock ``Demonstration of end-to-end automation of {DNA} data storage,''
\newblock {\em Scientific reports}, vol. 9, no. 1, pp. 1--5, 2019.

\bibitem{dimopoulou2019biologically}
Melpomeni Dimopoulou, Marc Antonini, Pascal Barbry, and Raja Appuswamy,
\newblock ``A biologically constrained encoding solution for long-term storage
  of images onto synthetic {DNA},''
\newblock in {\em 2019 27th {E}uropean {S}ignal {P}rocessing {C}onference
  ({EUSIPCO})}, 2019.

\bibitem{dimopoulou2021imageVQ}
Melpomeni Dimopoulou and Marc Antonini,
\newblock ``Image storage in {DNA} using {V}ector {Q}uantization,''
\newblock in {\em 2020 28th {E}uropean {S}ignal {P}rocessing {C}onference
  ({EUSIPCO})}. {IEEE}, 2021, pp. 516--520.

\bibitem{dimopoulou2021jpeg}
Melpomeni Dimopoulou, Eva Gil San~Antonio, and Marc Antonini,
\newblock ``A {JPEG}-based image coding solution for data storage on {DNA},''
\newblock in {\em 2021 29th {E}uropean {S}ignal {P}rocessing {C}onference
  ({EUSIPCO})}. {IEEE}, 2021, pp. 786--790.

\bibitem{pan2019image}
Chao Pan, SM~Yazdi, S~Kasra Tabatabaei, Alvaro~G Hernandez, Charles Schroeder,
  and Olgica Milenkovic,
\newblock ``Image processing in {DNA},''
\newblock {\em arXiv preprint arXiv:1910.10095}, 2019.

\bibitem{SFC4}
Xavier Pic and Marc Antonini,
\newblock ``A constrained shannon-fano entropy coder for image storage in
  synthetic dna,''
\newblock {\em European Signal Processing Conference (EUSIPCO)}, 2022.

\bibitem{secilmis2022towards}
Luka Secilmis, Michela Testolina, Davi Lazzarotto, and Touradj Ebrahimi,
\newblock ``Towards effective visual information storage on dna support,''
\newblock in {\em Applications of Digital Image Processing XLV}. SPIE, 2022,
  vol. 12226, pp. 29--35.

\bibitem{pic2023rotating}
Xavier Pic, Eva Gil~San Antonio, Melpomeni Dimopoulou, and Marc Antonini,
\newblock ``Rotating labeling of entropy coders for synthetic dna data
  storage,''
\newblock {\em IEEE International Conference on Digital Signal Processing
  (DSP)}, 2023.

\bibitem{DNAcoding}
Melpomeni Dimopoulou, Marc Antonini, Pascal Barbry, and Raja Appuswamy,
\newblock ``A biologically constrained encoding solution for long-term storage
  of images onto synthetic dna,''
\newblock {\em European Signal Processing Conference (EUSIPCO)}, 2019.

\end{thebibliography}

\end{document}